\documentstyle[12pt]{article}
\begin{document}
\title{\large \hspace{10cm} ITEP-PH-14/99 \\ \hspace{10cm} December 1999 \\
\vspace{1cm}
\LARGE \bf
Some comments on $\bar n p$-annihilation branching ratios
into $\pi \pi$-, $\bar K K$- and $\pi \eta$-channels}
\author {
A. E. KUDRYAVTSEV \thanks{E-mail: kudryavtsev@vitep5.itep.ru}
\\
{\it Institute of Theoretical and Experimental Physics,}\\
{\it B.Cheremushkinskaya, 25, Moscow, 117259, Russia}\\
}
\date{}
\maketitle
\vspace{1mm}
\centerline{\bf {Abstract}}
\vspace{3mm}

     We give some remarks on the $\bar n p$-partial branching ratios
in flight at low momenta of antineutron, measured by OBELIX
collaboration. The comparison is made to the known branching
ratios from the $p \bar p$-atomic states. The branching ratio
for the reaction $\bar n p \to \pi^+\pi^0$ is found to be
suppressed in comparison to what follows from the $ p \bar p$-data.
It is also shown, that there is no so called dynamic
$I=0$--amplitude suppression for the process $N\bar N \to K\bar K$.

\newpage

\section{Some useful definitions}

Let us consider first the $N\bar N$-system. By definition $\mid I,I_3>$ is
the isospin wave function
of the $N\bar N$-system with isospin $I$ and
its projection $I_3$.
Using notations of ref. \cite{1}, we write the following
relations between the physical states $\mid N\bar N>$ and
states of definite isospin $\mid I,I_3>$:

\begin{equation}
\mid p\bar p>={1\over\sqrt2}[\mid 1,0>-\mid0,0>],~~\mid n\bar n>={1\over\sqrt2}[\mid 1,0>+\mid0,0>].
\label{1}
\end{equation}

On the contrary in terms of physical states the wave function $\mid I,I_3>$
looks for isosinglet state as
\begin{equation}
\mid 0,0>=-{1\over\sqrt2}[\mid p\bar p>+\mid n\bar n>],
\label{2}
\end{equation}
and for isotriplet as
\begin{equation}
\mid 1,-1>=\mid n\bar p>,
\mid 1,0>={1\over\sqrt2}[\mid p\bar p>-\mid n\bar n>],
\mid1,1>=\mid \bar n p>.
\label{3}
\end{equation}
Each wave function is normalized as:
$$<N \bar N \mid N \bar N>=1,~~~ <I,I_3\mid I,I_3>=1. $$
Let us also define wave function for the hadron final state $\mid a>$
with definite isospin
$I$: $\mid a>_I$. We shall use the notations $\hat V^I_a$
for transition operator from initial $\mid I, I_3 >_{N\bar N}$-state 
to $\mid a>_I$ 
and
\begin{equation}
V^I_a=_I<a\mid \hat V^I_a \mid I,I_3>_{N\bar N},
\label{4}
\end{equation}
is martix element for this operator. It doesn't depend on $I_3$. 
Evidently  that
$$\hat V^I_a\mid J,J_3>_{N\bar N}=0$$
in the case $I\not=J$.

\section{Matrix elements for the transitions
$N\bar N\rightarrow \pi\pi$ and $N\bar N\to K \bar K$.}

Consider only the transitions to the final $\pi\pi$-states from
the initial $N\bar N$
$S$-wave
($^3S_1$ ). In this case the $\pi\pi$-system is produced in $I=1$ isospin
state. So there is only one operator $\hat V^1_{\pi}$.
The expansion of the $\mid\pi\pi>$-wave function in terms of the states
with definite isospin has the form:
\begin{equation}
\mid\pi^+\pi^->={1\over \sqrt3}\mid 0,0>
+{1\over \sqrt2}\mid 1,0>+{1\over \sqrt6}\mid 2,0>,
\label{5}
\end{equation}

$$\mid\pi^+\pi^0>={1\over \sqrt2}\mid1,1>-{1\over \sqrt2}\mid 2,1>.$$

Thus using definitions (1),
(3)
and (4), we get

\begin{equation}
<\pi^+\pi^0\mid \hat V^1_{\pi} \mid \bar n p>={1\over \sqrt2}V^1_{\pi},
\label{6}
\end{equation}

$$<\pi^+\pi^-\mid \hat V^1_{\pi}\mid p \bar p>={1\over 2} V^1_{\pi}.$$
It means, that the processes $p \bar p \to \pi^+\pi^-$ is to be at least
by factor two suppressed in
comparison to  $\bar n p\to \pi^+\pi^0$.

Let us now consider the transitions into $K\bar K$-final states.
Isospin wave functions for $K\bar K$-states have the following form:

\begin{equation}
\mid K^+K^->={1\over \sqrt2}[\mid 1,0>-\mid 0,0>],
\label{7}
\end{equation}

$$ \mid K^0 \bar K^0>=-{1\over \sqrt2}[\mid 1,0>+\mid 0,0>],$$
\begin{equation}
\mid K^+\bar K^0>=\mid 1,1>,~~~~\mid K^0K^->=-\mid 1,-1>.
\label{8}
\end{equation}

In this case $\mid K \bar K>$ ~final state is indeed a mixture of both
$I=0$ and $I=1$ isospin states ($I_3=0$).
Hence both operators $\hat V^1_K$ and $\hat V^0_K$ give
contribution to this reaction, and
$$_K<1,I_3\mid \hat V^1_K \mid 1,I_3>_{N\bar N}= V^1_K,
~~~~_K<0,0\mid \hat V^0_K \mid 0,0>_{N\bar N}=V^0_K.$$
In terms of
$V^1_K$ and $V^0_K$ we may calculate matrix elements between the physical
states:

\begin{equation}
<K^+K^- \mid \hat V_K \mid p \bar p>={V^0_K+V^1_K \over 2},
~~~~<K^0\bar K^0 \mid \hat V_K \mid p \bar p>={V^0_K-V^1_K \over 2},
\label{9}
\end{equation}
\begin{equation}
<K^+ \bar K^0 \mid \hat V_K \mid    \bar n p>= V^1_K.
\label{10}
\end{equation}

The matrix elements (9),(10) are related  to the corresponding 
partial cross-sections: 

$$\sigma=4\pi {q \over k} \mid <f\mid V \mid i>\mid^2 ,$$
where $q$ and $k$ are final and initial c.m. momenta.
We get the agreement for the expression (10) with what is
given in the ref.[1],
but
expressions (9)
differ from that of ref. \cite{1}.
Namely, redefining the operators according to equation (32)
of ref. [1], we get:
\begin{equation}
\sigma(p \bar p \to K^+K^-)+\sigma(p \bar p \to K^0 \bar K^0)=\mid A_0 \mid ^2
+\mid A_1 \mid ^2
\label{11}
\end{equation}
and
\begin{equation}
\sigma( \bar n p \to K^+ \bar K ^0 )=2 \mid A_1 \mid ^2.
\label{12}
\end{equation}
Notice, that factor 2 in the right-hand side of the equation (12) 
is not present  in
equation (35) of the
paper [1]. Historically this factor was also lost
in the papers
\cite{2,3}, and this error was reproduced later in some review
papers, see, e.g. \cite{4,5}. That is why the conclusion of the papers
\cite{1,2,3} on $I=0$-amplitude suppression seems to be incorrect
and is to be
revised. We shall discuss this problem in Section 4.

\section{Some relations between branching ratios
 in $p \bar p$-  and $\bar n p $-
annihilation processes}

Let us first consider the $\pi\pi$-case.
By definition of the branching ratio we have:
$$Br_{\pi^+\pi^0}(\bar n p)={\sigma(\bar n p \to \pi^+\pi^-) \over \sigma (
\bar n p
\to all)}$$
and similar expression for the $p \bar p$-case. So the 
ratio of branching ratios is:
\begin{equation}
{Br_{\pi^+\pi^0}( \bar n p ) \over Br_{\pi^+\pi^-}( \bar p p)}=
{\sigma (\bar n p \to \pi^+\pi^0) \over \sigma ( \bar n p \to all )}:
{\sigma (\bar p p \to \pi^+\pi^-) \over \sigma(\bar p p \to all)}.
\label{13}
\end{equation}

Notice, that at low energies, if only $S$-wave contribute, we have:
\begin{equation}
\sigma(p\bar n \to \pi^+\pi^0)=4 \pi {3 \over 4} \mid
<\pi\pi \mid \hat V^1_{\pi}
\mid p\bar n > \mid ^2 {q \over k}
\label{14}
\end{equation}
and
\begin{equation}
\sigma(p \bar p \to \pi^+\pi^-)=4 \pi {3 \over 4} C^2(k) \mid
< \pi\pi \mid
\hat V^1_{\pi}
\mid p \bar p > \mid ^2 {q \over k}.
\label{15}
\end{equation}
Here $C^2(k)$ is the Gamov factor,
$$C^2(k)={2\pi \over k a_B}/[1-\exp(-{2\pi \over ka_B})],$$
and $a_B=57.6 fm$~is the $p\bar p$-Bohr radius. 
Taking into account (13)-(15), we get:
\begin{equation}
{Br_{\pi^+\pi^0}(\bar n p ) \over Br_{\pi^+\pi^-}(\bar p p)}=
{\mid <\pi^+\pi^0 \mid \hat V^1_{\pi} \mid \bar n p > \mid ^2 \over
\mid < \pi^+\pi^- \mid \hat V^1_{\pi} \mid \bar p p> \mid ^2 }
{[\beta C^{-2}(k)\sigma^{ann}(p\bar p \to all)] \over
[\beta \sigma ^{ann}(\bar n p \to all)]}
\approx 2R,
\label{16}
\end{equation}
where $R$ is now a well defined and finite quantity:
\begin{equation}
R= {\lim_{k \to 0}[\beta C^{-2}(k)\sigma^{ann}(p \bar p)] \over
\lim_{k \to 0} [\beta \sigma^{ann}(\bar n p)]}.
\label{17}
\end{equation}
From the experimental data of refs. [6] and [8] we get the value of R at
low momenta of incident antiproton
($p_{lab}=50-70 MeV/c$):
\begin{equation}
R= {32\pm2 \over 25.3 \pm 1.0}\approx 1.26 \pm 0.10.
\label{18}
\end{equation}

Notice, that this value coincides with what follows from the experimental
data on annihilation
of antiprotons off deuteron \cite{7}. So we conclude, that the data \cite {8}
on total annihilation $\bar n p$-cross section are in agreement with
the results of quite independent experiment for
the annihilation  of antiproton on deuteron \cite{7}.
One may find the more detailed
discussion of value R extracted from the
different data on deuteron and  some heavier 
nuclei in the review paper \cite{9}.

A case of kaons looks very similar.
Using  eqs. (9)-(10)
as well as the definition of the ratio R (17), one gets the following
relation between branchings for the reactions $p\bar p \to K^+K^-$,
$p \bar p \to K^0 \bar K^0$ and $\bar n p \to K^+\bar K^0$:

\begin {equation}
{\mid V^1_K\mid ^2+ \mid V^0_K \mid ^2 \over 2\mid V^1_K \mid ^2 }=
R {Br(p \bar p \to K^+K^-)+Br(p\bar p \to K^0\bar K^0) \over
Br(\bar n p \to K^+\bar K^0)}.
\label{19}
\end{equation}

\section{The analysis of the experimental situation}
In Ref. \cite{8} the branching ratio 
for the reaction $\bar n p \to \pi^+\pi^0$
in the momentum interval 50-150 MeV/c (S-wave) was found to be equal:
\begin{equation}
Br( \bar n p\to\pi^+\pi^0)=(2.3 \pm0.4)10^{-3}.
\label{20}
\end{equation}
This value is to be compared with what follows from the $(p \bar p)$-atomic
experiment
for the reaction
$p\bar p \to \pi^+\pi^-$.
The separation of the S- and P-wave contribution to  last reaction was
provided in the Refs. \cite{10,11}. So we get for the branching ratio 
into $\pi^+\pi^-$-channel 
from atomic S-state:

~~~~~~~~~~~~~~~a)~~~~$(2.37\pm 0.23)10^{-3}$~~~[10];~~~~~
b)~~~~$(2.04\pm 0.17)10^{-3}$~~~[11].~~~~~~~~~~~~~~~~~~~~~

Substituting these numbers into eq.(16), we get the evident contradiction.
It means, that something is wrong with the branchings.
If one believes in the experimental branchings for both $\bar n p$-
and $\bar p p$-channels, the only possible way to solve the problem
is to suggest, that the $p \bar p$-atomic wave function at small distances
has an abnormal admixture of the $\bar n n$-component. We shall discuss this
hypothesis
in the next Section.

Now let us discuss a case of kaons. The
only information on branching ratio $N\bar N \to K\bar K$ for isospin
$I=1$ channel for long time was available from the old data
for absorption of antiproton on deuteron \cite{12},
$$Br(\bar p n\to K^0K^-)=(1.47\pm 0.21)10^{-3}.$$
Nowadays the OBELIX collaboration gives [1] ($S$-wave):
$$Br(\bar n p \to K^+K_S)=(0.92 \pm0.23)10^{-3}.$$
It means, that the branching into $K^+\bar K^0$ is:
$$ Br(\bar n p \to K^+\bar K^0)=2Br(\bar np\to K^+K_S)=(1.84\pm0.46)10^{-3}.$$
It is seen, that this last number for branching does not contradict
the old data by  Bettini et al. [12].

At the same time from the ASTERIX experiments [3,13] we have:
$$Br(p\bar p \to K^+K^-)=(1.08\pm 0.05)10^{-3},$$
$$Br(p\bar p \to K^0\bar K^0)=(0.83\pm0.05)10^{-3}.$$
Using these values and taking into account equation
(19), we get
\begin{equation}
\mid V^0_K\mid \approx 1.3 \mid V^1_K\mid.
\label{21}
\end{equation}
So we conclude, that there is no evidence for any suppression of
$I=0$--amplitude
for the reaction $N\bar N \to K\bar K$ in the S-wave. The dynamic selection
rule for this process, declared in the Refs.[1-5]  is the consequence of
 incorrect formulae for branchings  used in refs.[1,2].

Let us also discuss a case of  $\pi\eta$--channel. From the
experiment [8] it follows, that in the momentum interval~~150-250 MeV/c (P-wave)
$$Br(\bar n p\to \pi^+\eta)=(0.99\pm0.22)10^{-3}.$$
At the same time from the paper [10] we have:
$$Br(p\bar p \to \pi^0\eta)=(7.7\pm1.13)10^{-4}.$$
So again we come to the conclusion that the ratio
$${Br(\bar np\to \pi^+\eta) \over Br(\bar pp \to \pi^0\eta)}$$
is significantly less than $2R$ (see eq.(16)).

\section{Possible solution of the problem for the $N\bar N \to \pi\pi$
branchings}

In line with the papers [1,14,15]
let us suppose, that the wave function for $p\bar p$--atom at small distances
is a superposition of $\mid p\bar p>$ and $\mid n\bar n>$ configurations, i.e.:
\begin{equation}
\mid \psi_{at}>={1 \over \sqrt{1+\epsilon^2}}
[\mid p\bar p>+\epsilon \mid n\bar n>].
\label{22}
\end{equation}
In terms of the states of definite isospin it means, that
\begin{equation}
\mid \psi_{at}>={1 \over \sqrt{2(1+\epsilon^2)}}[(1-\epsilon)\mid 1,0>-
(1+\epsilon)\mid 0,0>].
\label{23}
\end{equation}
So
it follows immediately, that:
\begin{equation}
{Br(\psi_{at}\to \pi^+\pi^-) \over Br(\bar n p \to \pi^+\pi^0)}=
{(1-\epsilon)^2 \over 2(1+\epsilon^2)R}
\label{24}
\end{equation}
A case $\epsilon=0$ corresponds to the usual suggestion of the absence of
the
$n \bar n$-component in the $p\bar p$--atom. In the limit $\epsilon=-1$
the atomic state is that of definite isospin $I=1$. Substituting the
experimental numbers
for the
 $\pi\pi$-branchings  (see Section 4), we conclude, that
it is possible  to fit the parameter $\epsilon$
so the equation (23) is justified.
For example, taking Br$(p\bar p\to \pi^+\pi^-)=1.87$ (lower limit)
and Br$(\bar np\to \pi^+\pi^0)=2.7$ (upper limit), we get
$\epsilon=-2.24$, that corresponds to the value of mixing angle
$\cos \alpha= 1/ \sqrt {1+\epsilon^2};~~ \alpha \approx 66^{\circ}$.
It means, that the admixture of the $\bar n n$--component
 should be large to fit
the experimental data.

\section{Conclusion}

a) The data on the $\bar np$--total annihilation cross section, presented
by 
OBELIX Collaboration [8], are in agreement with the data
on the value of the ratio $R$, determined from the
absorption of
antiprotons on deuteron (see [7] and references in [9]).

b) The branching ratios for the reactions $\bar np \to \pi^+\pi^0$ and
$\bar n p\to \pi^+\eta$ at low energies [8] seem to be too large in
comparison to what follows from
the analysis of the known branching ratios for the $p\bar p$--atom.

c) The branching for the reaction $\bar np\to K^+K_S$ is in agreement
to the known branching for the reaction $\bar p n\to K^0K^-$ from the
deuteron
data [12]. There is no suppression for the $I=0~~ N\bar N\to K\bar K$--reaction
amplitude in S-wave (no specific dynamic selection rule).

d) Some admixture of the $\mid n \bar n>$-component in the
$p\bar p$-atomic wave function
may help in solving the problems with the branching into two pions
and $\pi\eta$. However to solve this problem, the admixture should be 
large enough.

\bigskip

\begin{center}
\large
\bf
ACKNOWLEDGMENTS
\end{center}

\bigskip

I am very thankful  to  G. Bendiscioli, B. Giacobbe, B.O.Kerbikov,
A. Rotondi, A.Zenoni and A. Zoccoli for useful discussion of preliminary
results of this research.

This work was supported in part by the Russian Foundation for Basic Research 
under grants 96-15-96578 and 98-02-17618.


\newpage

\begin{thebibliography}{100}
\bibitem{1} Jaenicke J., Kerbikov B., Pirner H.// Z.Phys. 1991. V.A339. P.297.
\bibitem{2} Landua R. "Proton-Antiproton Interactions and Fundamental
Symmetries" Proeuropean Symposium, Mainz, F.R.Germany,
5-10 Sept.1988;
Nucl.Phys.B(Proc.Suppl)1989. V.8. P.179.  
\bibitem{3} Doser M. et al.// Phys. Lett.B. 1988. V.215, P.792.
\bibitem{4} Amsler C. and Myhrer F.// Ann. Rev. Nucl. Sci. 1991. V.41, P.219.
\bibitem{5} Dover C., Gutsche T., Maruyama M. and Faessler A.//
 Prog. Part.
Nucl. Phys.1992. V.29. P.87.
\bibitem{6} Bertin A. et al.// Phys.Lett.B. 1996. V.369, P.77.
\bibitem{7} Bizzarri R.et al.// Nuovo Cim.A.1974. V.22, P.225 ;
 Kalogeropulos T. et al.// Phys. Rev.D. 1980. V.22, P.2585;
Riedelberger J.et al.// Phys. Rev.C. 1989. V.40, P.2717.
\bibitem{8} Giacobbe B.(OBELIX Collaboration). The talk, given at LEAP'96
Conference, Dinkelsbuhl, Germany, August 27-31,1996; A.Bertin et al.,
Nucl. Phys.B (Proc. Suppl.) 1997. V.56A. P.227.
\bibitem{9} Bendiscioli G. and Kharzeev D.// La Rivista del Nuovo Cim.
1994. V.17. I.6. P1.
 \bibitem{10} Peters K. "Low Energy Antiproton Physics". Proc. of the
 LEAP'94 Conf., Bled, Slovenia, Sept 12-17, 1994, p.3.
World Scientific. Singapure-New Jersey-London-Hong Kong.
\bibitem{11} Batty C.J.// Nucl. Phys.A. 1996. V.601, P.425.
\bibitem{12} Bettini A.  et al. //Nuovo Cim.A. 1969. V.62, P.1038.
\bibitem{13} Doser M. et al. //Nucl.Phys.A. 1988. V.486, P.493.
\bibitem{14} Carbonell J., Ihle G., Richard J.M.// Z.Phys.A. 1989. V.334, P.329.
\bibitem{15} Klempt E.// Phys.Lett.B.1990. V.244, P.122.
\end{thebibliography}
\end{document}